\PassOptionsToPackage{pdfpagelabels=false}{hyperref} 
\documentclass[fleqn, usenatbib,useAMS,amsmath, twocolumn]{mnras}
\usepackage{mathtools, xspace}

\newcommand\lsim{\mathrel{\rlap{\lower4pt\hbox{\hskip1pt$\sim$}}
        \raise1pt\hbox{$<$}}}
\newcommand\gsim{\mathrel{\rlap{\lower4pt\hbox{\hskip1pt$\sim$}}
        \raise1pt\hbox{$>$}}}

\usepackage{times}
\usepackage{graphicx}
\usepackage{verbatim}
\usepackage{tikz}
\usepackage{color}

\usepackage[T1]{fontenc}
\usepackage{aecompl}
\usepackage{calligra}
\DeclareMathAlphabet{\mathcalligra}{T1}{calligra}{m}{n}
\DeclareFontShape{T1}{calligra}{m}{n}{<->s*[2.2]callig15}{}

\usepackage{hyperref}
\hypersetup{
    pdfnewwindow=true,      % links in new window
    colorlinks=true,       % false: boxed links; true: colored links
    linkcolor=blue,          % color of internal links
    citecolor=orange,        % color of links to bibliography
    filecolor=blue,      % color of file links
    urlcolor=blue           % color of external links
}

\def\prd{PRD}
\def\apj{ApJ}                 % Astrophysical Journal
\def\apjl{ApJL}               % Astrophysical Journal, Letters
\def\mnras{MNRAS}             % Monthly Notices of the RAS
\def\aap{A\&A}                % Astronomy and Astrophysics
\def\nat{Nature}              % Nature
\def\obs{\rm{obs}}

\def\min{\rm{min}}
\def\max{\rm{max}}
\def\med{\rm{med}}

\def\Msun{{M_{\odot}}}
\def\Rsun{{r_{\odot}}}

\def\crit{\rm{crit}}

\def\dynesty{{\tt Dynesty}\xspace}

\def\BHcirc{\tikz\draw[black,fill=black] (0,0) circle (.25ex);}
\def\spin{\mathcal{S}}
\def\IBCO{\rm{ibco}}

\begin{document}

\title[Constraining the Stellar Mass Function]{Constraining the Stellar Mass Function from the Deficiency of Tidal Disruption Flares in the Nuclei of Massive Galaxies}
\author[D. J. D'Orazio, A. Loeb, J. Guillochon]{Daniel J. D'Orazio$^1$\thanks{daniel.dorazio@cfa.harvard.edu},
  Abraham Loeb$^1$, James Guillochon$^1$ \\
     $^1$Astronomy Department, Harvard University, 60 Garden Street, Cambridge, MA 02138}

\maketitle
\begin{abstract}
The rate of tidal disruption flares (TDFs) per mass of the disrupting black
hole encodes information on the present-day mass function (PDMF) of stars in
the clusters surrounding super massive black holes. We explore how the shape
of the TDF rate with black hole mass can constrain the PDMF, with only weak dependence
on black hole spin.  We show that existing data can marginally constrain the
minimum and maximum masses of stars in the cluster, and the high-mass end of
the PDMF slope, as well as the overall TDF rate. With $\mathcal{O}(100)$ TDFs
expected to be identified with the Zwicky Transient  Facility, the overall
rate can be highly constrained, but still with only marginal constraints on
the PDMF. However, if $\lsim 10 \%$ of the TDFs expected to be found by LSST
over a decade ($\mathcal{O}(10^3)$ TDFs) are identified, then precise and
accurate estimates can be made for the minimum stellar mass (within a
factor of two) and the average slope of the high-mass PDMF (to within
$\mathcal{O}(10\%)$) in nuclear star clusters. This technique could be adapted
in the future to probe, in addition to the PDMF, the local black hole mass
function and possibly the massive black hole binary population.
\end{abstract}

\begin{keywords} % commenting results in missing "thanks notes"!
stars: luminosity function, mass function -- galaxies: nuclei
\end{keywords}

\maketitle

\section{Introduction}

A star can be ripped apart by a supermassive black hole (BH) when the tidal
force across the star exceeds the gravitational force that binds the star.
Depending on the relative sizes and masses of the star and BH, this can occur
before, or after the star plunges across the BH event horizon. If disruption
occurs outside of the event horizon, we may observe the event as a tidal
disruption flare \citep[TDF, \textit{e.g.},][]{Hills:1975, Rees:1988,
Gezari+2006, Gezari+2009, vanVelzen+2011, Bloom+2011, Gezari+2012,
Chornock+2014, Arcavi+2014, Holoien+2014, Vinko+2015, Tadhunter+2017}, else
the event is likely dark or significantly different in appearance,
\citep[\textit{e.g.}][]{,DL:2013,DL:2016,Lu:2017a}.

Recently \citet[][hereafter V18]{vanVelzen:2018} estimated the rate of TDFs
per mass of the disrupting BH and showed a steep drop in TDF rate above
$M_{\BHcirc} \sim 10^8 \Msun$, approximately the BH mass above which the
disruption of a $1 \Msun$ star is hidden from observation behind the BH
horizon. Because the measured BH mass function does not drop off as
steeply as this TDF mass function, V18 takes this finding as evidence for a
horizon, and the validity of the TDF hypothesis for events identified as such.

Here we show that this fall-off not only demonstrates the existence of a
horizon, but also encodes rich information on the present day stellar mass
function (PDMF) in the nuclear star clusters surrounding super massive BHs.
This is because the properties of the star that is being disrupted directly
affects the properties of the resultant TDF, and thus the population of TDFs
represents the underlying population of stars that is being disrupted.

We use the volumetric rate of TDFs to constrain the mass function of stars in
the nuclear clusters surrounding super massive BHs. This is determined by the shape of the rate
distribution with BH mass, not its overall normalization, which depends on the
number of stars prone to disruption in addition to the details of stellar
dynamics \citep{MagTrem:1999, WangMerritt:2004, StoneMetzger:2016}.
Specifically, we exploit the relation between the tidal radius scale and the
BH event horizon scale to predict the minimum stellar mass that can be
disrupted outside the horizon of a BH of specified mass. Because the PDMF sets
the number of stars above a certain mass, the TDF rate per BH mass encodes
the shape of the PDMF.

Using a sample of $12$ TDFs with BH mass measurements from V18, and
generating a new sample of 17 TDFs by combining the V18 dataset with the
dataset from \citet[][hereafter M18]{MocklerJames+2018}, we marginally
constrain the low mass and high mass cutoffs for stars in the PDMF,
$M_{\min} = 0.93^{+5.7}_{-0.85} \Msun$ and $M_{\max} = 43.0^{+105.0}_{-33.0}
\Msun$; an average metallicity of stars in the nuclear cluster,
$\log_{10}(Z/Z_{\odot})=0.94^{+0.88}_{-0.94}$; and using a broken power for
the PDMF, we constrain the high-mass PDMF slope of $\xi = -1.6^{+1.0}_{-2.0}$,
with very little sensitivity to the BH spin. The PDMF and the above parameters
are defined below in Eq. (\ref{Eq:MR}).

While the existing data does not provide significant constraints on the PDMF,
we find that the precision on at least the minimum stellar mass and the slope
of the PDMF will improve by an order of magnitude in the LSST era, when $\gsim
10^3$ identified TDFs will be available. Hence, the novel technique presented
here would prove powerful in constraining the PDMF within galactic nuclei,
where intriguing evidence indicates top-heavy IMFs
\citep[\textit{e.g.},][]{Bartko_GCSMF+2010} and high metallicities
\citep[\textit{e.g.},][]{TDo+2018}.

In \S~\ref{S2} we present our analytic calculation of the conditional
probability for measuring a specific BH mass given the observation of a TDF.
In \S~\ref{S3} we present constraints on the PDMF from present and expected
future observations of the BH-mass dependent TDF rate. In \S~\ref{S4} we
discuss possible improvements to this model and conclude.

\section{Methods}
\label{S2}

We calculate the probability that a galaxy's central super massive BH has a mass
$M_{\BHcirc}$ given that a TDF is observed in that galaxy. This probability
can be written via Bayes' theorem as,
\begin{equation}
\mathcal{P} \equiv  P(  M_{\BHcirc} | \rm{TDF} ) = \frac{ P( \rm{TDF} | M_{\BHcirc}) P(M_{\BHcirc}) }{P(\rm{TDF})}.
\label{Eq:Bayes}
\end{equation}

We are interested in the shape of this probability distribution as a function of
$M_{\BHcirc}$. With this in mind, we assume that $P(\rm{TDF})$ is a constant
that encodes the overall normalization of the rate at which stars are
delivered onto orbits that will be captured by super massive BHs and result in a TDF. This
term could have a weak dependence on BH mass \citep{Brockamp+2011,
WangMerritt:2004, Kochanek:2016, StoneMetzger:2016}, a feature we plan to explore
in future work. For $P(M_{\BHcirc})$, we choose the local BH mass function of
\citet{Shankar+2004}. Hence, we are left with calculating the likelihood term
in the numerator of Eq. (\ref{Eq:Bayes}): the conditional probability that a
star will be disrupted, given knowledge of the BH mass.

\subsection{Probability of a TDF given the BH mass} 

A star with mass $M_*$ and radius $r_*$ will be disrupted if it passes within
a distance \citep{Hills:1975},
\begin{equation}
r_T = \left(\frac{M_{\BHcirc}}{M_*}\right)^{1/3} r_*,
\label{Eq:TidRad}
\end{equation}
of a black hole of mass $M_{\BHcirc}$. If the BH mass is too large, or the
star is too tightly bound, the disruption will occur inside the event horizon
of the BH. No TDF will be observed when $r_T \geq r_{\IBCO}$, where, 
\begin{equation}
r_{\IBCO} = \frac{GM_{\BHcirc}}{c^2} \left[ 2 \mp \spin + 2 \sqrt{1\mp\spin} \right],
\end{equation} 
is the inner-most-bound circular orbit (IBCO) of the BH
in the equatorial plane \citep{Bardeen72, LevinGabe:2009}, which depends on the BH mass and spin, $\spin$.\footnote{We are not
aware of any analytic expression for the general angle dependent IBCO for arbitrary spin
\citep{Hod:2017, Will:2012}.} Inside of this radius the star will plunge into the BH.

Because the value of $r_{\IBCO}$ is different for prograde and retrograde
stellar orbits (ranging from $GM/c^2$ for $\spin=1$ to $(3+2\sqrt{2})GM/c^2$ for
$\spin=-1$), we assume that there is no preference for stellar angular
momentum relative to BH spin and take the average value,
\begin{equation}
\left< r_{\IBCO} \right>_{\spin} = \frac{GM}{c^2}\left[2 + \sqrt{1-\spin} + \sqrt{1+\spin} \right].
\label{Eq:<rIBCO>}
\end{equation}
This quantity is always larger than the horizon radius and has a smaller spin
dependent range than the horizon radius. This is a consequence of including
retrograde orbits which will cause the star to be captured at a much larger
radius without disruption. For a single interaction, $r_{\IBCO}$ changes the
radius at which a star will be swallowed before disruption by a large amount,
a factor of almost $6$ between maximally spinning prograde and retrograde
orbits. When averaging over many interactions, however, the spin of the BH
does not have a large effect, statistically, on the disruption of an infalling
star. It will, however, contribute to an intrinsic scatter in the rate.

Note that, for simplicity, we average the spin dependent value of the
IBCO only over equatorial prograde and retrograde orbits. This choice yields
the maximal spin dependence of the IBCO. This is because the spin dependence
of the IBCO becomes weaker for orbits inclined away from the equatorial plane
\citep{Will:2012}. That is, averaging over the entire sphere of encounters
would result in an averaged IBCO with an even weaker dependence on the spin
than found in Eq. (\ref{Eq:<rIBCO>}), with a range more tightly
constrained to $4GM/c^2$.

To solve for the critical stellar mass, $M^*_{\rm{crit}}$, above which
disruption by a BH of given mass may occur, we must adopt a stellar 
mass-radius relation. Here we provide a simple fitting function to the main-sequence,
metallicity-dependent  mass-radius relation of \cite{Tout+1996} that matches
well observational data for metallicities $Z/Z_{\odot} \in (0.03, 3)$,
\begin{eqnarray}
&\log_{10}\left[\frac{r_*(M_*, Z)}{\Rsun} \right] = \min\left\{ \log_{10} \left[ 0.85 \left( \frac{M_*}{\Msun} \right)^{0.85} \right], 0 \right\} \nonumber \\
&+  \max\left\{  \log_{10} \left[ \left(\frac{Z}{Z_{\odot}}\right)^{0.096} \left( \frac{M_*}{\Msun} \right)^{0.60} \right], 0 \right\}. 
\label{Eq:r*Z}
\end{eqnarray}
Assuming $Z=Z_{\odot}$, we can solve analytically for the critical mass (see
Appendix). Otherwise we find $M^*_{\rm{crit}}$ by numerically solving
expression (\ref{Eq:TidRad}) for $M_*$ when $r_T = \left< r_{\IBCO}
\right>_{\spin}$ and with $r_*$ given by Eq. (\ref{Eq:r*Z}).

The probability that a star, on a trajectory towards the BH tidal disruption
radius, will disrupt outside of the event horizon is the same as the
probability that this star is above the critical mass $M^*_{\rm{crit}}$. This
probability can be derived from a model of the PDMF. We assume a PDMF
proportional to a Kroupa initial-stellar-mass function (IMF) with a
parameterized slope above $0.5 \Msun$,
\begin{eqnarray}
    \frac{dN_*}{dM_*} &\propto&  H(M_* - M_{\min}) [1 - H(M_* - M_{\max})] \times \nonumber \\
   &&\times \begin{cases} 
    C_1  \left(\frac{M_*}{\Msun}\right)^{-0.3} &  M_* < 0.08\Msun, \nonumber \\\\
    C_2 \left(\frac{M_*}{\Msun}\right)^{-1.3} & 0.08\Msun \leq M_* \leq 0.5 \Msun ,\nonumber \\ \\
    \left( \frac{M_*}{\Msun}\right)^{\xi} & M_* > 0.5\Msun, \nonumber 
    \end{cases} \nonumber \\
    C_1 &=& (0.5 \Msun)^{\xi + 1.3}, \qquad C_2 = \frac{C_1}{0.08 \Msun},
\label{Eq:MR}
\end{eqnarray}
and with lower and upper stellar mass cutoffs $M_{\min}$ and $M_{\max}$, and where
the constants $C_1$ and $C_2$ ensure that the power law breaks occur at the
correct value of $M_*$. $H$ is the Heaviside function.

The probability of a disruption, given the BH mass, is then,
\begin{equation}
P(\rm{TDF} | M_{\BHcirc}) \equiv P(M_* \geq M_{\rm{crit}} | M_{\BHcirc}) = \frac{ \int^{\infty}_{M^*_{\rm{crit}}}{ \frac{dN_*}{dM_*} dM_* } }{\int^{\infty}_{0}{ \frac{dN_*}{dM_*} dM_* }}.
\label{Eq:Pdef}
\end{equation}

Combining (\ref{Eq:r*Z})--(\ref{Eq:Pdef}), we find an analytic expression for
$P( \rm{TDF} | M_{\BHcirc})$ in terms of the numerically derived value of
$M_{\rm{crit}}$ (for a simplified analytical form of $M_{\rm{crit}}$, see the
Appendix). This allows us to compute the desired quantity, $P( M_{\BHcirc} |
\rm{TDF} )$ in Eq. (\ref{Eq:Bayes}). Because we are only interested in the
shape of $P( M_{\BHcirc} | \rm{TDF} )$ with $M_{\BHcirc}$, we choose
$P(\rm{TDF})=1/K_0$ by which the overall rate is normalized.

\subsection{Parameter dependencies} 
\label{S:paramdeps}

\begin{figure*}
\begin{center}$
\begin{array}{c}
\includegraphics[scale=0.5]{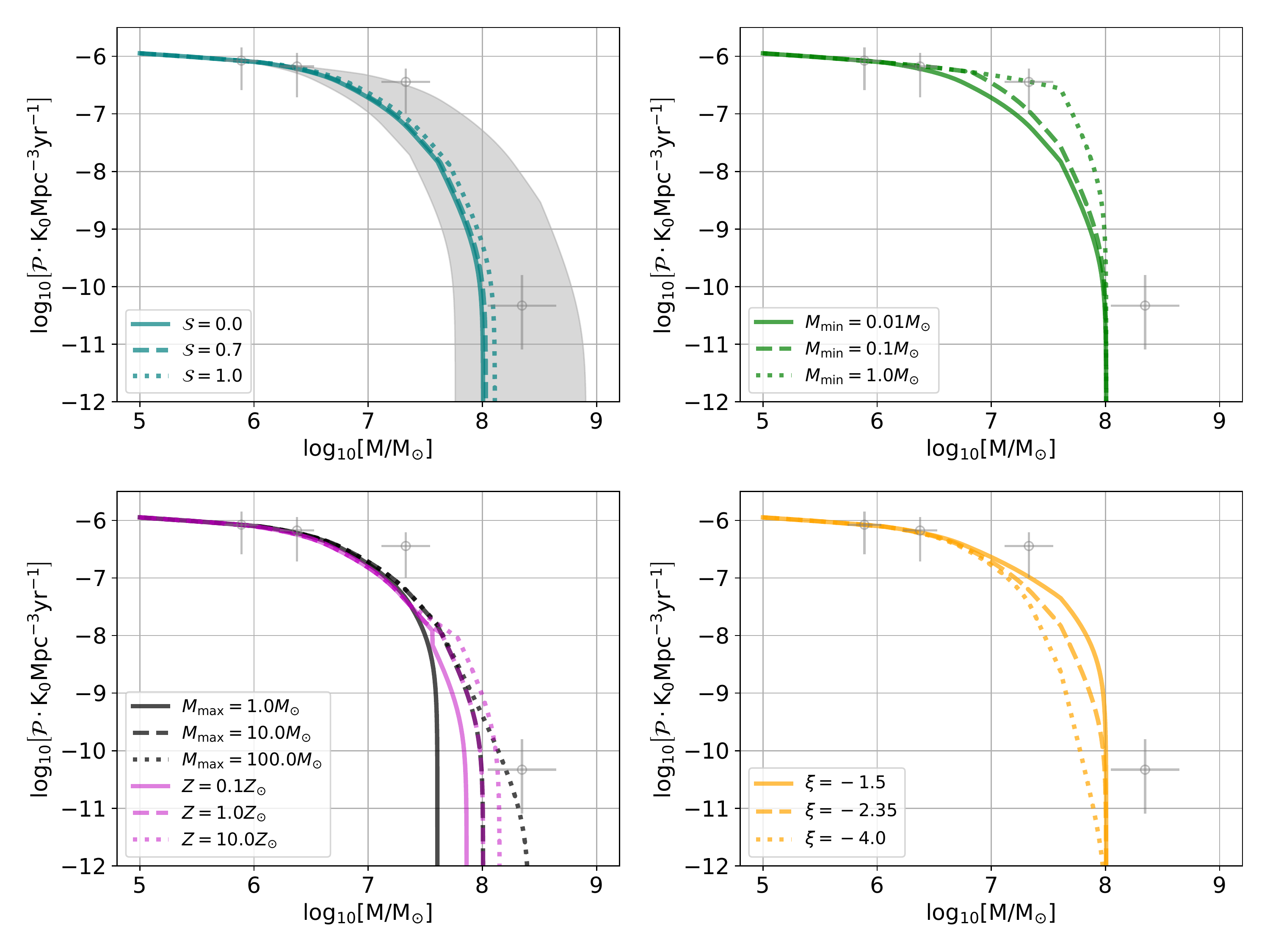}
\end{array}$
\end{center}
%\vspace{0pt}
%
\caption{
Model parameter dependencies. From top left to bottom right, the varied
parameters are the dimensionless BH spin, $\spin$; the minimum stellar mass,
$M_{\min}$; the maximum stellar mass, $M_{\max}$, (black) and metallicity, $Z$
(purple); and the high-mass slope of the PDMF $\xi$. The grey shaded region in
the top left panel is the range of values taken if the spin \textit{direction}
is not averaged. The grey data points are from \citet{vanVelzen:2018} for reference.
}
% %
\label{Fig:1}
\end{figure*}

We plot the derived TDF rate $P( M_{\BHcirc} | \rm{TDF} )$ in Figure
\ref{Fig:1} for various BH spins, stellar metallicities, and PDMFs, varying
the cutoff masses $M_{\min}$ and $M_{\max}$ as well as the high-mass
slope of the PDMF. We use fiducial parameter values of $(M_{\min}, M_{\max},
\spin, \xi, Z) = (0.01 \Msun, 10\Msun, 0, -2.35, Z_{\odot})$ unless otherwise
stated in the figure legends. As a reference we plot in grey the data for the
TDF rate from V18.

All of the realizations of the BH mass dependent TDF rate in Figure
\ref{Fig:1} show the same general behavior. For lower BH masses, there is a
shallow decrease in the TDF rate with BH mass that follows the BH mass
function $P(M_{\BHcirc})$. For higher BH masses,  there is a steep cutoff in
the rate. The cutoff marks where stars drawn from the prescribed PDMF are no
longer favored to be above the critical mass for disruption. The main effect
of changing the PDMF and BH spin is to change the shape and location of this
cutoff.

The top left panel of Figure~\ref{Fig:1} shows the change in the TDF rate as a
function of BH spin. The teal lines depict the TDF rate when considering
direction-averaged stellar orbits relative to the BH spin. This amounts to
using the spin-direction averaged value of $r_{\IBCO}$ presented in Eq.
(\ref{Eq:<rIBCO>}), valid for averaging many TDFs. The result implies
very little spin dependence with a small preference for higher spin BHs to
allow higher disruption rates at higher BH masses (associated with a shift of
the cutoff to higher BH masses).

For reference, the grey shaded region in the top left panel shows the range of
TDF rates possible for all values of the spin, \textit{i.e.}, the bounds of
the shaded region correspond to the rates expected for purely retrograde or
purely prograde stellar orbits around a maximally spinning BH. This indicates that
a large scatter in the BH spin dependent TDF rate is expected; however, for a
large number of TDFs, the average rate will converge to a result with very
little spin dependence. This scatter is a tracer of the range of BH spin
magnitudes in the sample.

The left panel also shows that a change in the spin, even if not averaging
over spin direction, affects the TDF rate uniformly with BH mass, shifting the
curve horizontally for different values of the spin. As we now see from the
remaining panels in Figure~\ref{Fig:1}, this is a very different change in the
rate than the shape change that results from varying the PDMF parameter values.

The bottom left panel of Figure~\ref{Fig:1} shows the change in the TDF rate
as a function of the maximum stellar mass, $M_{\max}$. Increasing the maximum
allowed stellar mass extends the TDF rate to higher values at high BH
masses. This is because higher mass main-sequence stars are less
tightly bound (Eq. \ref{Eq:r*Z}) and can thus be disrupted by higher mass BHs
(see Eq. \ref{Eq:TidRad}).

The purple lines in the bottom left panel of Figure~\ref{Fig:1} show the
effect of changing stellar metallicity, which has a similar effect as changing
the maximum mass. Higher metallicity stars above $1\Msun$ are less tightly
bound \citep[see][and Eq. \ref{Eq:r*Z}]{Tout+1996}. Hence, higher metallicity
stars can be more easily disrupted at higher BH masses. We consider
metallicity as high as those presented in Figure~\ref{Fig:1} because there is
evidence for high metallicity stellar populations (up to $10Z_{\odot}$) in
galactic nuclear star clusters \citep{TDo+2018}. When fitting to data,
however, we use more conservative constraints on $Z$ for consistency with the
metallicity with which the mass-radius relation is calibrated.

The top right panel of Figure~\ref{Fig:1} shows the change in the TDF rate as
a function of the minimum stellar mass, $M_{\min}$. Increasing $M_{\min}$, for
a fixed $M_{\max}$, narrows the range of possible stellar masses and hence
steepens the TDF rate cutoff at high BH mass. The minimum stellar mass does
not change the value of the maximum disrupting BH mass, as is the case for
$M_{\max}$. Instead, the value of $M_{\min}$ crucially changes the shape of
the cutoff by changing the range of stellar masses available for disruption;
the fraction of stars with mass above the critical mass for disruption drops
quickly when $M_{\min}$ and $M_{\max}$ are close in value, while the cutoff is
more gradual for disparate values of $M_{\min}$ and $M_{\max}$.

The bottom right panel of Figure~\ref{Fig:1} shows the change in the TDF rate
as a function of the high-mass slope of the PDMF, $\xi$. Steeper values of the
high-mass PDMF slope imply a lower fraction of high-mass stars and hence a
lower probability of disruption at higher BH masses. A steeper slope also
boosts the relative number of low mass stars. As a result, the effect of a
steeper PDMF slope on the mass dependent TDF rate is similar to that found for
decreasing $M_{\min}$ in the top left panel of Figure~\ref{Fig:1}.

In summary, the maximum stellar mass and the metallicity ($M_{\max}$, $Z$) set the BH
mass above which no TDFs occur. The minimum stellar mass and the high-mass PDMF slope
($M_{\min}$, $\xi$) set the shape of the cutoff. The direction-averaged BH spin
has very little influence on the BH mass dependent TDF rate, but could encode
the range of spin magnitudes in the scatter of TDF rates.

%%%%%%%%%%%%%%%%%%%%%%%%%%%%%%%%%%%%%%%%%%%%%%%%
%%% FIGURE %%%
%%%%%%%%%%%%%%%%%%%%%%%%%%%%%%%%%%%%%%%%%%%%%%%%    

\section{Application to present and future data}
\label{S3}

We fit our model to two data sets: the data in Figure~3 of V18
(available in the online version) and to an extended data set that
we have constructed using BH masses measured directly from the TDF light
curve (M18). We then fit our model to mock data that
could be gathered in the ZTF and LSST eras.

\subsection{van Velzen and Mockler data sets}
\label{S:vvMdata}

The construction of the V18 TDF volumetric rate per BH mass is presented in
V18. To construct the M18 data set, we used the maximum volume method
outlined in V18. For the BH masses, we use the union of the van Velzen and
M18 BH masses, using the mass from M18 if the TDF is represented in both
sets. For the quantity $z_{\max}$ we use only the value derived from the
survey flux limit and the peak TDF flux (namely, the value in Table 1 of V18).

We use both data sets and our model to constrain the mass function of stars in
galactic nuclei. To this goal we construct the likelihood that our model
$\mathcal{P}(M_{\BHcirc}; \mathbf{\Theta})$ with parameters $\mathbf{\Theta}$
is the correct model given the data $(p_i,M_{\BHcirc,i})$. Because the data is
not constructed from a simple binning across BH masses, but rather is
constructed with uncertainties in both coordinates we construct a log-likelihood
function that can take into account errors in both coordinates. We
generalize the one-dimensional $\chi^2$ to higher dimensions, which cannot be
done uniquely because of a choice for how to compute the distance between a
model and a data point. We choose to compute the minimum distance between point
and model. This results in choosing the line that orthogonally intersects the
model. 
This log likelihood is written
\begin{eqnarray}
-\ln \mathcal{L} &=& \sum_i
\frac{ \left(    \log_{10}p_i - \log_{10} \left[ \mathcal{P}_i(M^{\rm{proj}}_{\BHcirc,i}; \mathbf{\Theta}) \right]  \right)^2 }{2 \sigma^2_{ \log_{10}p, i } }  \nonumber \\
&+&  \frac{ \left(  \log_{10}M^{\obs}_{\BHcirc, i} -   \log_{10}M^{\rm{proj}}_{\BHcirc,i} \right)^2 }{2 \sigma^2_{
\log_{10}M_{\BHcirc},i } }   ,
\end{eqnarray}
where $M^{\rm{proj}}_i$ is the solution to the minimization of the distance
between data point and model for a given set of model parameters,
\textit{i.e.}, the x-coordinate for which a line between the data point and
the model is the shortest. Summation is over the $i^{\rm{th}}$ data point. The
quantities $\sigma^2_{ \log_{10} p_i }$ and $\sigma^2_{ \log_{10}M_{\BHcirc},i
}$ are the one-standard deviation uncertainties on the $i^{\rm{th}}$ data
point. In addition to the parameters, $\mathbf{\Theta} = (M_{\min}, M_{\max},
\spin, \xi, Z)$, we also vary the normalization constant $K_0$ with units of
Mpc$^{-3}$ yr$^{-1}$.

\begin{table}
    \caption{The ranges of the uniform priors used for nested sampling. Here $K_*$ is a constant that equals $10^{-4.125}$ for the V18 and M18 runs, and equals the true value of the normalization for the Mock-data runs.}
    \begin{center}
        \begin{tabular}{ c c }
        Parameter   & Bounds \\

        \hline
        $\log_{10}\left[  M_{\min}/\Msun \right]$ & $[-2.0, 2.0]$ \\
        $\log_{10}\left[  M_{\max}/\Msun \right]$ & $[-2.5, 2.5]$ \\
        $\spin$                                   & $[0.0, 1.0]$ \\
        $\xi$                                     & $[-5.0, 0.0]$ \\
        $\log_{10}\left[  Z/Z_{\odot} \right]$    & $[-2.45, 0.45]$ \\
        $\log_{10}\left[  K_0 \right]$            & $K_* \pm 1.0$
    \end{tabular}
    \end{center}
    \label{Table:Priors}
\end{table}

We use the dynamic nested sampling code \dynesty (Spiegel \textit{et al.}
\textit{in prep.})\footnote{http://dynesty.readthedocs.io/en/latest/}, based
on the algorithm of \citet{Higson:2017}, to sample the posterior
distribution. We assume uniform priors given in Table \ref{Table:Priors}. For
both the V18 and M18 data sets we present the median posterior values along with the
parameter range that falls within the $16^{\rm{th}}$ and $84^{\rm{th}}$
quantiles in Table $\ref{Table:BFpd}$.

\begin{table}
    \caption{The median-likelihood model parameter values, $\Theta\left(\mathcal{L}_{\med}\right)$, given the TDF rate per BH
    mass data from Figure~3 of \citet{vanVelzen:2018} and the data generated in this work from a combined sample of BH masses from \citet{MocklerJames+2018} and \citet{vanVelzen:2018}. Parameter uncertainties are
    quoted at $16^{\rm{th}}$ and $84^{\rm{th}}$ percentile values corresponding to
    $1\sigma$ uncertainties for Gaussian errors.}
    \begin{center}
        \begin{tabular}{ c c c }
        \hline
        \hline
            Parameter 

            & \multicolumn{2}{|c|}{ [$\Theta\left(\mathcal{L}_{\med}\right)^{+16\%}_{-84\%}$] } \\

            & V18 Best Fit Values (Fig. \ref{Fig:2vV})

            & M18 Best Fit Values  (Fig. \ref{Fig:3M})

            \\

        \hline
        $M_{\min}/\Msun$                         & $1.1^{+10.4}_{-1.0}$     & $0.93^{+5.7}_{-0.85}$  \\ \\
        $M_{\max}/\Msun$                         & $47^{+108}_{-37}$          & $43^{+105}_{-33}$ \\ \\
        $\spin$                                  & $--$       & $--$  \\ \\
        $\xi$                                    & $-2.3^{+1.6}_{-1.7}$      & $-1.6^{+1.0}_{-2.0}$ \\ \\
        $\log_{10}\left(Z/Z_{\odot}\right)$      & $-0.88^{+0.88}_{-0.96}$      & $-0.94^{+0.88}_{-0.94}$ \\ \\
        $\log_{10}K_0$                           & ${-4.12}^{+0.14}_{-0.14}$ & ${-4.45}^{+0.12}_{-0.11}$  
    \end{tabular}
    \end{center}
    \label{Table:BFpd}
\end{table}

\begin{table}
    \caption{The same as Table \ref{Table:BFpd}, but for the two sets of mock data displayed in Figure~\ref{Fig:4LSST}. For convenience, the mock data is generated with a value of the normalization $K_0$ that is different than in the case of the real data.}
    \begin{center}
        \begin{tabular}{ c c c c }
        \hline
        \hline
            Parameter 
            &
            & \multicolumn{2}{|c|}{ [$\Theta\left(\mathcal{L}_{\med}\right)^{+16\%}_{-84\%}$]  } \\

            & Truth Values

            & $N=400$ 

            & $N=4 \times 10^3$ \\

        \hline
        $M_{\min}/\Msun$                         &  $1.0$  &  $3.2^{+15.0}_{-2.8}$  & $1.7^{+1.6}_{-0.6}$ \\ \\
        $M_{\max}/\Msun$                         &  $70.0$  &  $65^{+109}_{-46}$  & $155^{+176}_{-86}$ \\ \\
        $\spin$                                  &  $0.7$  &  $--$  &  $--$ \\ \\
        $\xi$                                    &  $-2.0$ &  $-1.3^{+0.9}_{-2.0}$   & $-2.0^{+0.3}_{-0.3}$  \\ \\
        $\log_{10}\left(Z/Z_{\odot}\right)$      & $0.0$  &  $-0.8^{+0.8}_{-1.0}$   & $-0.6^{+0.6}_{-0.9}$ \\ \\
        $\log_{10}K_0$                           & $-1.60$ & ${-1.60}^{+0.02}_{-0.02}$  & ${-1.60}^{+0.01}_{-0.01}$ \\
    \end{tabular}
    \end{center}
    \label{Table:BFLSST}
\end{table}

\begin{figure*}
\begin{center}$
\begin{array}{cc}
\includegraphics[scale=0.49]{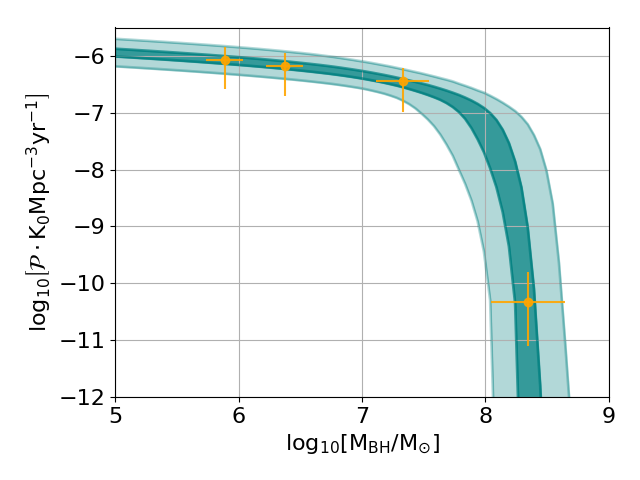}
\includegraphics[scale=0.21]{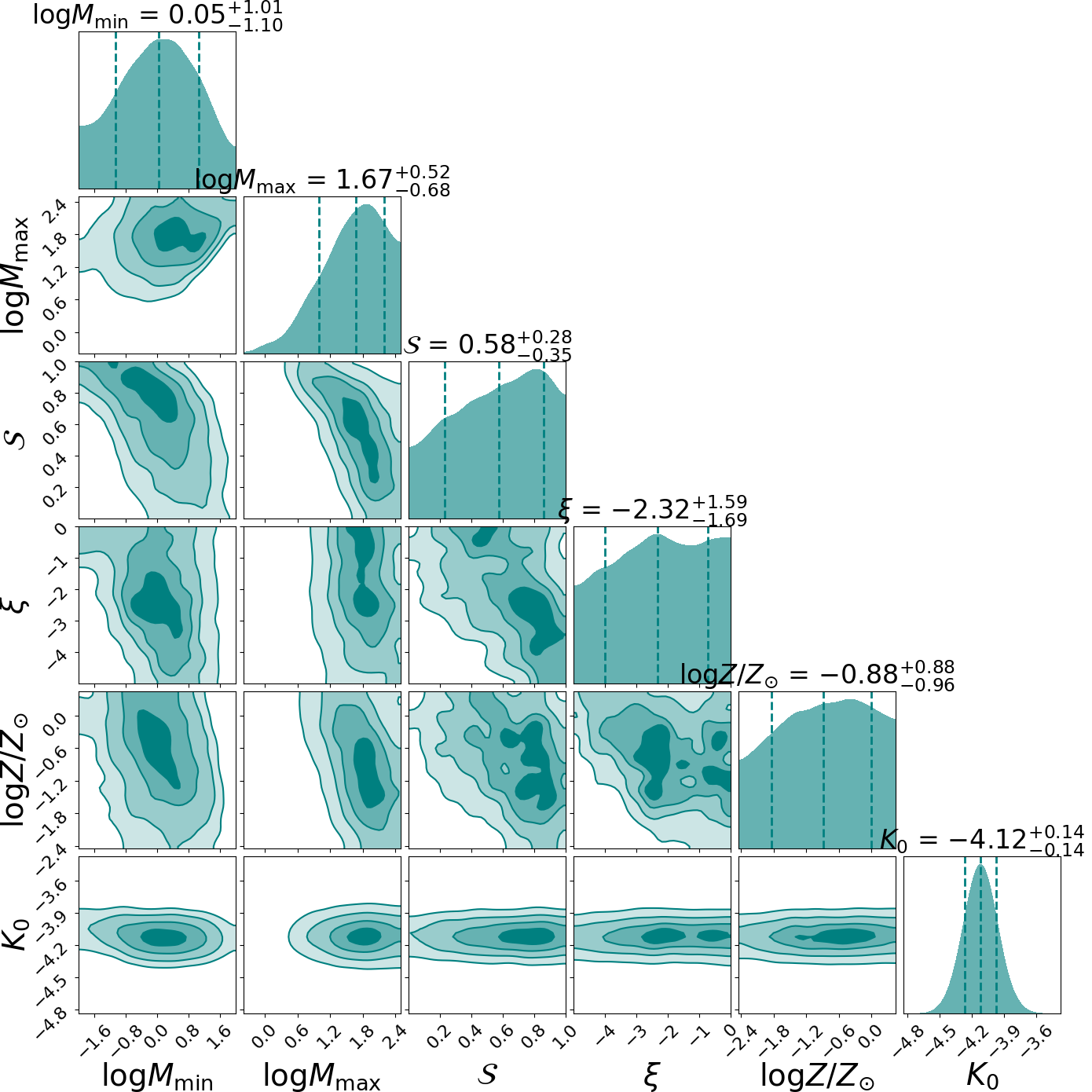}
\end{array}$
\end{center}
%\vspace{0pt}
%
\caption{\textit{Left:}  Orange points with error bars are the derived
volumetric rates of TDFs taken from \citet{vanVelzen:2018}. The shaded dark
(light) teal region represents the $16^{\rm{th}}$-$84^{\rm{th}}$ quantile
($5^{\rm{th}}$-$95^{\rm{th}}$ quantile) range of the highest likelihood models, Eq.
(\ref{Eq:Pdef}). \textit{Right:} Corner plot showing 2D and 1D representations
of the posterior distribution. The vertical lines on each 1D posterior
distribution represent the $16^{\rm{th}}$, $50^{\rm{th}}$, and $84^{\rm{th}}$
quantiles from left to right. With the small number of TDFs in this present
data set, the PDMF is not well constrained, however, these constraints will
improve greatly in the LSST era, as shown in Figure~\ref{Fig:4LSST}. }
% %
\label{Fig:2vV}
\end{figure*}

\begin{figure*}
\begin{center}$
\begin{array}{cc}
\includegraphics[scale=0.49]{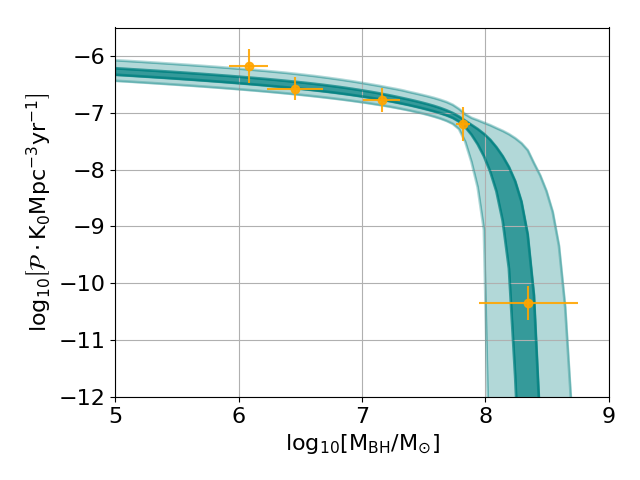}
\includegraphics[scale=0.21]{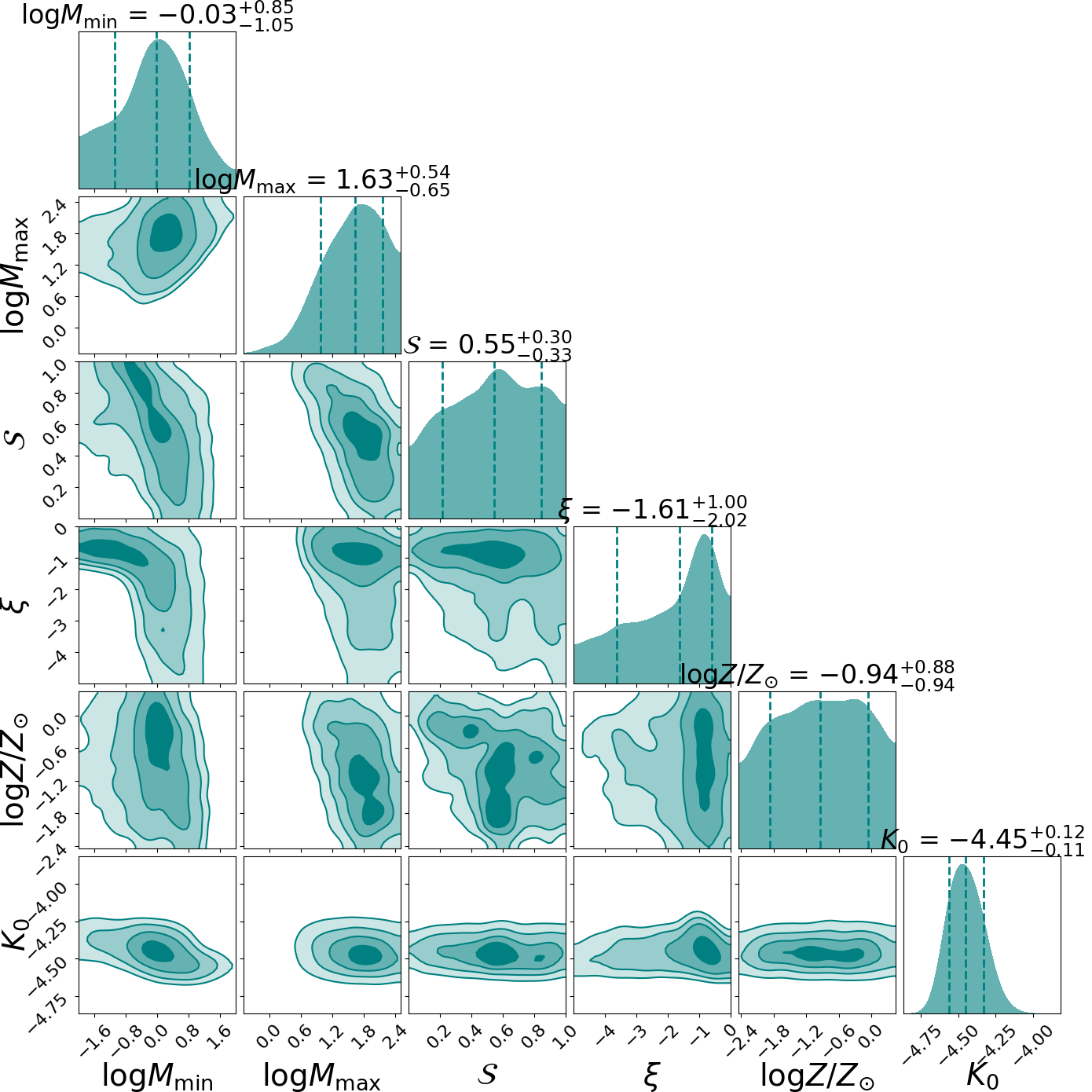}
\end{array}$
\end{center}
%\vspace{0pt}
%
\caption{
The same as Figure~\ref{Fig:2vV}, but for the data set derived from the union of the
\citet{vanVelzen:2018} and \citet{MocklerJames+2018} data sets as described in
\S~\ref{S:vvMdata}. The numbers of TDFs in the five bins are [3,8,4,1,1]. From the small number of TDFs in existing data, the PDMF is not well constrained, however, these constraints will improve greatly in the LSST era, as shown in Figure~\ref{Fig:4LSST}.
}
% %
\label{Fig:3M}
\end{figure*}

We find similar results for both data sets. The best fit parameter values are
displayed in Table \ref{Table:BFpd}. The results of fitting to the V18 data
are displayed in Figure~\ref{Fig:2vV} while the results from fitting to the
M18 data are displayed in Figure~\ref{Fig:3M}. Each figure shows the
$16^{\rm{th}}$-$84^{\rm{th}}$-quantile (dark teal) and
$5^{\rm{th}}$-$95^{\rm{th}}$-quantile (light teal) ranges of highest likelihood models
to the data (left panel) and corner plot representations of the 1D and 2D
posteriors (right panel). Note that the posteriors are not all Gaussian. This
is apparent in the location of the $16^{\rm{th}}$ and $84^{\rm{th}}$ quantiles
of each 1D posterior distribution denoted by the vertical dashed lines in the
corner plots.

While the V18 and M18 data sets are sparse, the data points at the highest BH masses
define the position and shape of the rate cutoff sufficiently to place a wide,
but not flat, constraint on the model parameter values. In the text below we quote
numbers from the M18 analysis, though the V18 numbers are similar.

\begin{itemize}
\item $\mathbf{M_{\min}}$: On the high end, the minimum stellar mass is not
significantly constrained. The median likelihood value is $0.9 \Msun$ and only
$16\%$ of the likelihood favors $M_{\rm{min}} \leq 0.1 \Msun$, possibly
suggesting preference for a minimum stellar mass $\gsim 0.1\Msun$.
\item $\mathbf{M_{\max}}$:  The median likelihood value of the maximum stellar mass
is $43 \Msun$ and ranges from $10 \Msun$ up to $148 \Msun$, for the $16\%$ and
$84\%$ quantiles. Any value in this range should be taken as a lower limit for
the maximum stellar mass. This is because $M_{\max}$ can only truly be probed
if the observed disruptions require such a high-mass star. These disruptions
are rare, the rarity of which depends on the PDMF slope $\xi$. The more
TDFs observed, the higher the chance that high-stellar-mass
disruptions could be found. Hence the lower limit on $M_{\max}$ will
increased with additional TDFs until its true value will be tightly constrained.
\item $\boldsymbol{\mathcal{S}}$: 
The BH spin is not constrained. The is because we use the retrograde/prograde
averaged value of $r_{\IBCO}$ to denote the point of no return for an
infalling star (Eq. \ref{Eq:<rIBCO>}). As shown in Figure~\ref{Fig:1}, this
results in very little dependence of our model on the BH spin. As
noted below Eq. (\ref{Eq:<rIBCO>}), we expect this insensitivity to be robust
as our expression for the average radius of the IBCO is a conservative choice,
allowing maximum spin dependence. Note that the fall-off in posterior
probability (right panels of Figures \ref{Fig:2vV} and \ref{Fig:3M}) for the
spin at high and low values, is due to smoothing of the posterior, resulting
in an artificial drop in probability at the prior boundaries. Hence the
labeled constraints $\mathcal{S}=0.5^{+0.3}_{-0.3}$ in Figures \ref{Fig:2vV}
and \ref{Fig:3M} are artificial. In the next section we show that there is
still no spin dependence even when considering a sample of $4 \times 10^3$
TDFs.

We note, however, that the spin averaging that we invoke is only valid when
each BH mass bin contains many more than a single TDF. As the V18 and M18 data sets have bins
with $1$ or $2$ TDFs, this assumption breaks down. One can think of these
small number bins as requiring larger y-error bars than shown due to a random
draw of BH spin in the range $[-1,1]$.
\item$\mathbf{\xi}$: The PDMF slope above $0.5 \Msun$ is not well constrained but
prefers shallow, top-heavy values, pushing up against the prior limit of
$\xi=0$. 
\item $\mathbf{Z/Z_{\odot}}$: The average metallicity of the stellar population is
consistent with solar, but also prefers high values, and like the PDMF slope,
pushes up against the prior value of $3 Z_{\odot}$. 
\end{itemize}

From the corner plots in both Figures \ref{Fig:2vV} and \ref{Fig:3M}, one can
see that there is a degeneracy in $M_{\min}$ and $\xi$. This degeneracy is
already suggested by Figure~\ref{Fig:1}; both $M_{\min}$ and $\xi$ affect the
shape of the rate cutoff at intermediate BH masses. This degeneracy can,
however, be removed in future data sets with more data for TDFs caused by BHs
in the mass range $\log_{10}(M/\Msun) \sim 7.5-8$.

The main differences in the results from the V18 and the M18 data sets is the
reduced degeneracy between $M_{\min}$ and $\xi$. This is because of the extra
TDF directly at the knee of the TDF-rate cutoff in the M18 data. This leads to
an even higher estimate for $M_{\min}$, and a slightly lower estimate for
$\xi$ (Table \ref{Table:BFpd}). Recall, however, that this difference is
due to only one TDF.

In the next section we show definitively that more data points at higher BH
masses do indeed remove the $M_{\min}$-$\xi$ degeneracy.

%%%%%%%%%%%%%%%%%%%%%%%%%%%%%%%%%%%%%%%%%%%%%%%%
%%% FIGURE %%%
%%%%%%%%%%%%%%%%%%%%%%%%%%%%%%%%%%%%%%%%%%%%%%%%    

%%%%%%%%%%%%%%%%%%%%%%%%%%%%%%%%%%%%%%%%%%%%%%%%
%%% FIGURE %%%
%%%%%%%%%%%%%%%%%%%%%%%%%%%%%%%%%%%%%%%%%%%%%%%%    

\subsection{Constraints from the ZTF and LSST}

The Zwicky Transient Facility (ZTF), which has recently been commissioned, could
detect $10-70$ TDFs per year \citep{SiftSaph:2017}. The Large Synoptic Survey
Telescope \citep[LSST;][]{LSST} is expected to discover $4 \times 10^3$ TDFs per year
\citep{vanVelzen+2011}, starting in the 2020s. Because the BH mass can be
measured directly from the TDF lightcurve itself (M18), it will be possible to
build a TDF BH mass function for hundreds of events in the ZTF era and
thousands of events in the LSST era. While identifying TDFs is difficult,
methods involving photometric cuts and spectroscopic follow-up, when feasible, have
proven effective \citep[\textit{e.g.}][]{SiftSaph:2017}. Here we create mock
samples of TDFs and associated BH masses based on an optimistic ZTF
scenario/pessimistic LSST scenario (400 TDFs) and an expected LSST scenario
($4 \times 10^3$ TDFs).

For each mock data set, we choose a truth model and randomly sample $N$ draws
from this distribution. We bin these draws into $\sqrt{N}$ evenly sized mass
bins with $y$ error given by the $\sqrt{N}$ Poisson estimate. To generate the
probability distribution from which we draw, we normalize Eq. (\ref{Eq:Pdef})
by the integral over Eq. (\ref{Eq:Pdef}) from $10^5 \Msun$ to infinity.

To fit to the binned data, we use a Poisson likelihood,
\begin{equation}
-\ln\mathcal{L} = \sum_i \log{N_i!} + \mu_i - N_i \log{\mu_i} ,
\end{equation}
where $N_i$ is the number of counts in the $i^{\rm{th}}$ mass bin, and the
model expectation value for the number of counts in a bin of size $\Delta M$ is
$\mu_i \equiv N_{\rm{tot}} \tilde{\mathcal{P}}_i \Delta M$, given the
normalized probability $\tilde{\mathcal{P}}_i$ at the center of the mass bin
and the total number of events $N_{\rm{tot}}$. In practice we approximate
$\log{N!} \approx N\log{N}-N +1/2\log{\pi(2N + 1/3)}$, which agrees with the
exact expression to at least the third decimal place for all $N$.  We again
use the \dynesty code to sample the posterior distribution and recover the best
fit parameter values and error bars. The assumed priors are listed in Table
\ref{Table:Priors}.

The two scenarios considered here represent an optimistic scenario at the end
of ZTF and the beginning of LSST ($N=400$), and an expected LSST scenario
($N=4 \times 10^3$). For 10 years of an LSST-like survey, these correspond to
$1\%$ and $10\%$ of all transients classified as TDFs purely from their
lightcurves.

\begin{figure*}
\begin{center}$
\begin{array}{cc}
\includegraphics[scale=0.5]{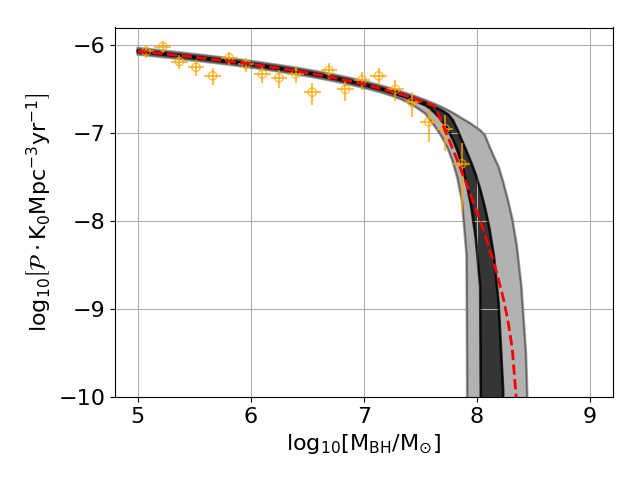}&
\includegraphics[scale=0.22]{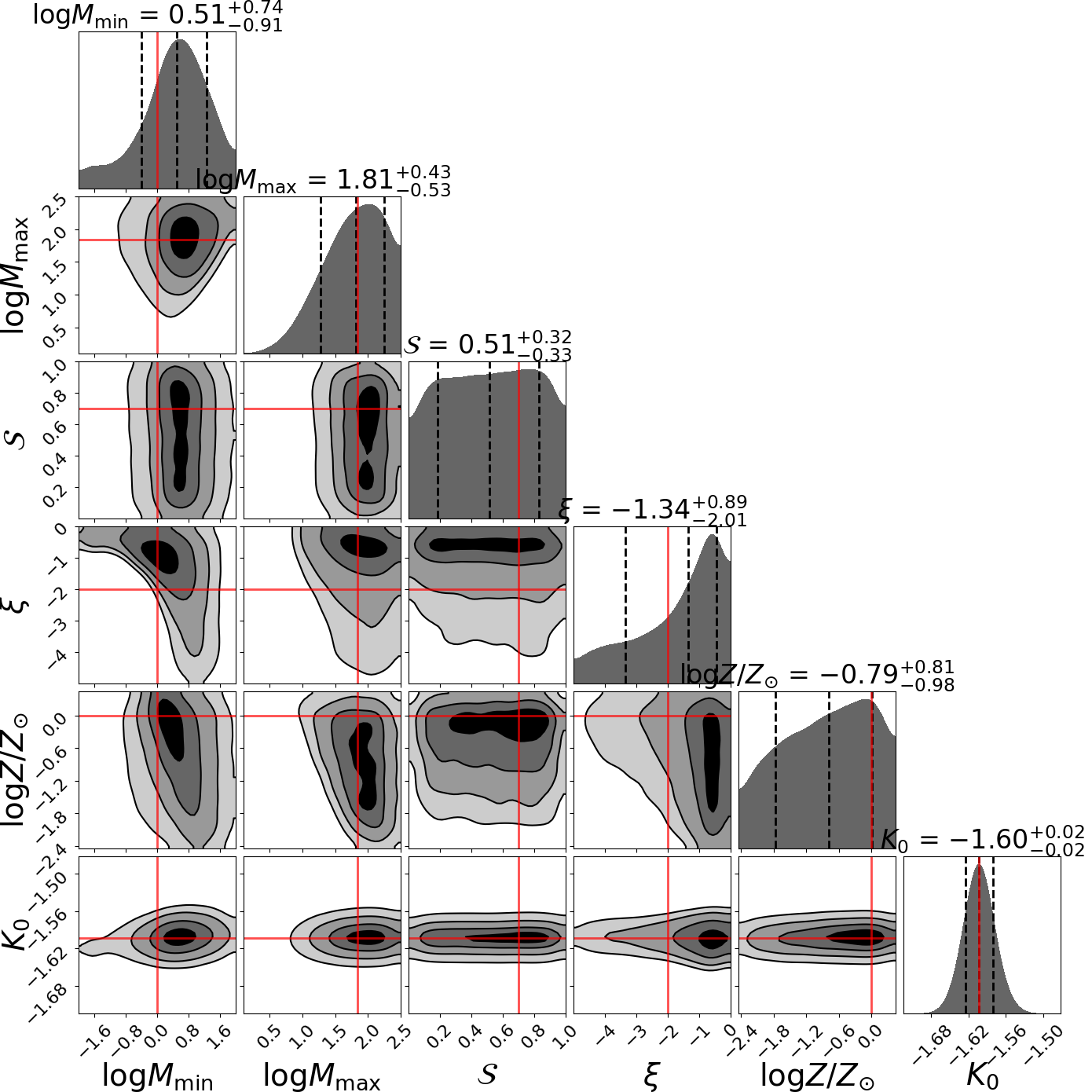} \\
\includegraphics[scale=0.5]{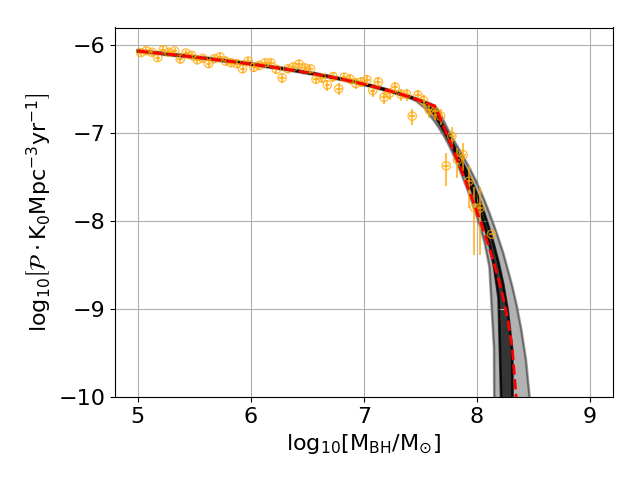}&
\includegraphics[scale=0.22]{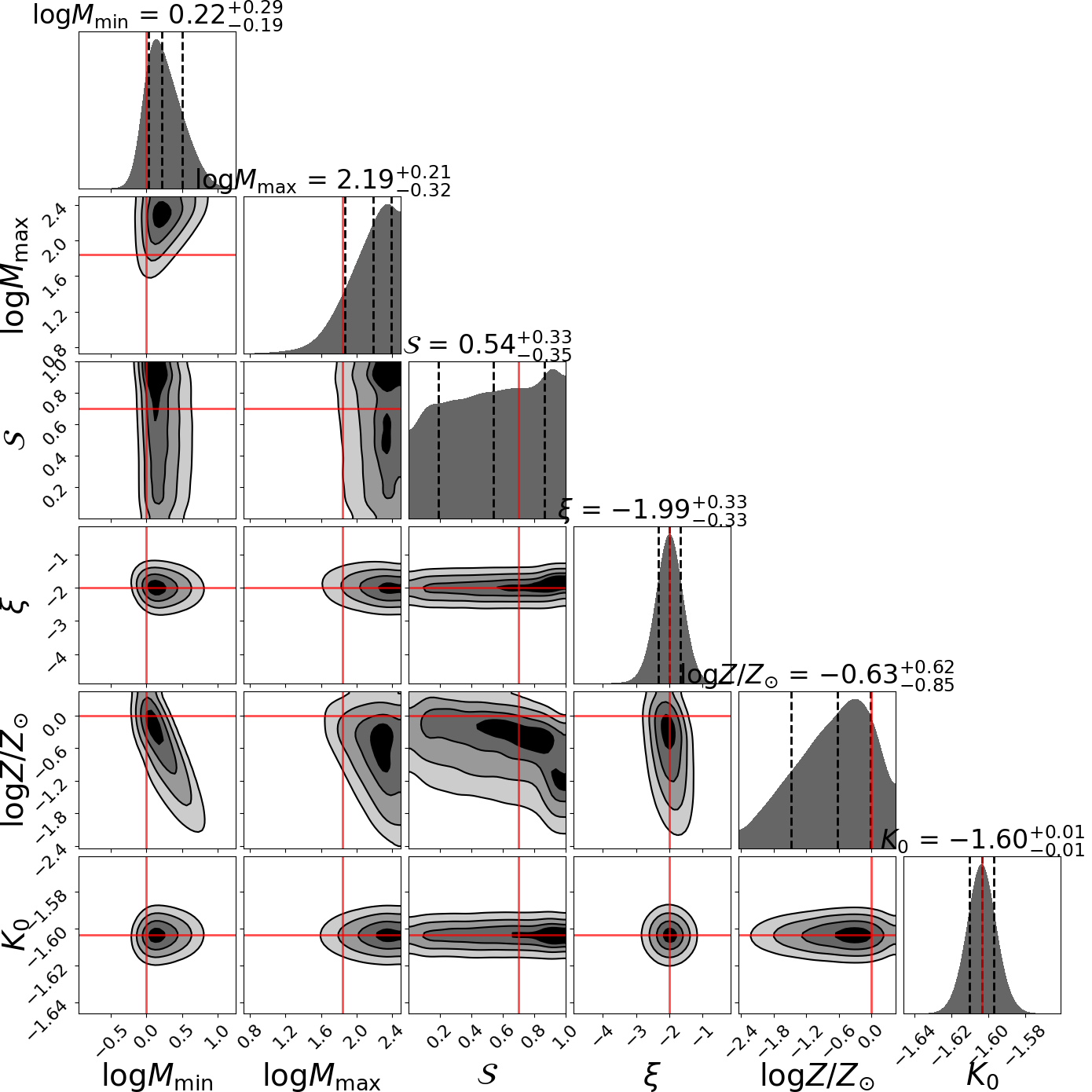} 
\end{array}$
\end{center}
\caption{ 
The same as Figures \ref{Fig:2vV} and \ref{Fig:3M}, but for mock data for 400 TDFs (top row) and $4 \times 10^3$ TDFs (bottom row), inspired by
the number of TDFs that could be identified by the ZTF and by the yearly rate expected for LSST respectively. The assumed parameter values
(truth values), from which the data are drawn are indicated by red lines in the right panels. The true model is shown as a dashed red line in the left panels.
}
% %
\label{Fig:4LSST}
\end{figure*}

The best fit parameter values and the underlying truth values are tabulated in Table
\ref{Table:BFLSST}. The results of fitting to both mock data sets are
displayed in Figure~\ref{Fig:4LSST}. The top row displays results for the ZTF
($N=400$) scenario and the bottom row displays results for the LSST ($N=4
\times 10^3$) scenario. The right column of Figure~\ref{Fig:4LSST} shows the
1D and 2D posteriors from a combined ensemble of 10 different random draws
from the underlying truth distribution, while the left column shows one of
these realizations (orange data points) as well as the true underlying model
(dashed red) and the $16^{\rm{th}}$-$84^{\rm{th}}$ quantile and
$5^{\rm{th}}$-$95^{\rm{th}}$ quantile uncertainties (black and grey shaded
regions) from the parameter recovery. In the right panels, the truth values
are indicated by red lines. We find that the truth values can be recovered
well for both data sets.

In both scenarios, the normalization parameter $K_0$ is recovered very well
simply because most of the TDFs fall at low BH mass where the normalization is
set. As in the real data sets, the BH spin parameter is again unconstrained
for both the ZTF and the LSST scenarios.

We find only a moderate enhancement in parameter constraints from the present
day data to the ZTF ($N=400$) scenario For example, the ZTF mock data exhibits
the same degeneracy between $M_{\min}$ and $\xi$ and has only slightly smaller
$16^{\rm{th}}$-$84^{\rm{th}}$ quantile error bars (see Tables \ref{Table:BFpd}
and \ref{Table:BFLSST}). In the ZTF scenario, however, the overall rate
normalization is now precisely constrained to within $\mathcal{O}(1\%)$ as
opposed to $\mathcal{O}(10\%)$ with the present day data.

Moving to the LSST scenario, we find a large improvement in parameter recovery accuracy
and constraints for the $M_{\min}$ and $\xi$ parameter values. The constraint on
the metallicity is only moderately improved, while the constraint on the maximum
mass is largely unchanged from the ZTF scenario.

For $M_{\min}$ and $\xi$, the $16^{\rm{th}}$-$84^{\rm{th}}$-quantile error
bars are reduced by almost an order of magnitude. More interestingly the
degeneracy between the two is removed and the PDMF slope $\xi$ is recovered
accurately and precisely. While $M_{\min}$ is recovered more precisely, and
still consistently, it is overestimated in both the ZTF and LSST scenarios.

The large improvement in parameter recovery between the ZTF ($N=400$) and LSST
($N=4 \times 10^3$) scenarios suggests that $N\gsim 10^3$ is the target number of TDFs
needed to make meaningful constraints on the PDMF, especially the PDMF slope
$\xi$ and the minimum stellar mass $M_{\min}$. An even greater number of TDFs
are needed to probe the maximum stellar mass if it is of order the fiducial
value of $70\Msun$ chosen here. This is expected, however, because the rarest
events, involving the disruption of the most massive stars by the most massive
BHs, constrain the maximum stellar mass. We discuss this further in the next
section.

%%%%%%%%%%%%%%%%%%%%%%%%%%%%%%%%%%%%%%%%%%%%%%%%
%%% FIGURE %%%
%%%%%%%%%%%%%%%%%%%%%%%%%%%%%%%%%%%%%%%%%%%%%%%%    

\section{Discussion and Conclusions}
\label{S4}

We have shown that the BH mass dependent TDF rate can provide constraints on
the PDMF of stars in galactic nuclei. In particular, one can constrain the
minimum and maximum stellar mass, the high-mass slope of the
PDMF, and the average stellar metallicity. For a large enough sample of TDFs,
the BH spin enters through the scatter in the TDF mass dependent rate, an
analysis of which we leave for the future.

Fitting to existing data, consisting of 17 TDFs for which a BH mass can be
estimated, provides a preliminary demonstration of the constraining power of
this method. Our best fit parameter values and $16^{\rm{th}}$-$84^{\rm{th}}$-quantile 
uncertainties are recorded in Table \ref{Table:BFpd}. While the
uncertainties on the best fit parameter values are large, we find reasonable values
for the PDMF model parameters.

Fitting to mock data relevant for the end of the ZTF era shows that the
overall TDF rate can be constrained to within $\mathcal{O}(1\%)$, however, the
PDMF is still only marginally constrained with $400$ TDFs (see the top row of
Figure~\ref{Fig:4LSST} and Table~\ref{Table:BFLSST}).

Fitting to mock data relevant for the LSST era, consisting of $4 \times 10^3$
confirmed TDFs, which corresponds to identification of only $10\%$ of expected
TDFs over a $10$ year LSST lifetime, we find that the minimum stellar mass in
the cluster can be constrained within a factor of two for an assumed $1\Msun$
truth value, and that the high-mass PDMF slope can be constrained to of order
$\pm0.3$. A lower limit can be placed on the maximum stellar mass in the
cluster (see the bottom row of Figure \ref{Fig:4LSST} and Table
\ref{Table:BFLSST}).

Most of the constraining power originates from observations of TDFs
with high-mass BHs. As discussed in \S \ref{S:paramdeps}, this is because the
rate of TDFs generated by low mass BHs reflect only the BH mass function. TDFs
generated by BHs approaching a few times $10^7\Msun$ hold information on the
minimum stellar mass and the PDMF slope. TDFs generated by $\sim10^8 \Msun$
BHs determine the maximum stellar masses and the metallicity. However, the
high-information content TDFs from the highest mass BHs are more rare and
hence, statistically, require a larger sample of TDFs to gather. This is the
main reason why a larger sample of TDFs is required to constrain the PDMF.

We note, however, that when generating mock data, we do not assume any
dependence of TDF luminosity on stellar or BH mass. If, however, TDFs
disrupted by more massive BHs, for example, are intrinsically brighter, then
the volume within which they can be observed is larger compared to TDFs
associated with less massive BHs. The result being that our mock data
under-samples the number of TDFs that would be associated with the more massive
BHs. Hence, our estimate of the constraining power of the TDF rate is
conservative. Note that this is only an issue when generating mock data,
the real data of course take this into account via the maximum-volume method
used in v18.

This point is indeed hinted at by the existence of the ASASSN-15lh TDF. This
event defines the high BH mass cutoff in the TDF rate and influences the fits
to the PDMF model displayed in Figures \ref{Fig:2vV} and \ref{Fig:3M}. Within
the model, however, such an event is highly improbable and such events only
begin to appear in the LSST mock data. For the fiducial PDMF parameter values,
and assuming uniform intrinsic luminosity across the BH mass spectrum, the
ASASSN-15lh event is an approximately one in $10^4$ event. However, because
ASASSN-15lh is $\sim \times 10$ brighter than other TDFs, the volume of
observation is increased by a factor of $10^3$, making ASASSN-15lh more like a
one in 10 event.

Note that, the ASASSN-15lh event, while important for determining the
cutoff in TDF rate at high BH mass, mostly provides information only on the
maximal mass for the analysis of the V18 and M18 data. Removing the 
ASASSN-15lh data point from the analysis, we find that the recovered parameter values
and uncertainties are not significantly altered. The largest excursion is for
the maximum stellar mass estimate, decreasing in median value and increasing
in range to $25^{+120}_{-22} \Msun$ ($32^{+127}_{-27} \Msun$ for the M18 data)
as opposed to $47^{+108}_{-37} \Msun$ ($43^{+105}_{-33} \Msun$ for the M18
data) when ASASSN-15lh is included. The minimum stellar mass estimate changes
to $0.6^{+7.5}_{-0.5} \Msun$ ($1.0^{+5.3}_{-0.9} \Msun$ for the M18 data)
compared to $1.1^{+10.4}_{-1.0} \Msun$ ($0.93^{+5.7}_{-0.85} \Msun$ for the
M18 data) when ASASSN-15lh is included. The other parameters stay nearly the
same with only small increases in the $16^{\rm{th}}$-$84^{\rm{th}}$-quantile
uncertainties. This suggests that ASASSN-15lh may be more important for
signifying a cutoff in the rate and motivating the TDF model than it is for
constraining the PDMF, at least at the level of detail present in the V18 and
M18 data sets. We suspect that when more data in the knee of the differential
TDF rate function is populated, the inclusion of ASASSN-15lh could have a
larger effect on the accuracy of the recovered parameters.

The probability and influence of an ASASSN-15lh event aside, we note
that one should be cautious when considering only a single TDF at the high BH
mass end of the TDF rate function. It could be that ASASSN-15lh is not due to
the tidal disruption of a star at all \citep[\textit{e.g.},][]{Dong15lh+2016,
Metzger+2015, SukhWoos+2016, Margutti+2017, Kruhler+2018}. That possibility
aside, even if ASASSN-15lh is a bona-fide TDF, its location in the existing
data could be skewed a few different ways. One possibility is that this event
is the result of a highly spinning BH \citep[\textit{e.g.},][]{Leloudas+2016}.
Because the highest BH mass bin consists only of the single ASASSN-15lh data
point, we could be prone to a larger uncertainty than reflected by the
Poissonian error bars. Future additions to the TDF rate will determine the
extent to which ASASSN-15lh is an outlier.

Another interesting possibility is that ASASSN-15lh is the product of a
disruption by a compact, super massive BH binary with a disparate mass ratio
\citep{Coughlin15lh+2018}, thereby allowing a large central BH mass
measurement, yet disruption by a much smaller BH. The possibility of this
scenario depends on the unknown probability of finding a super massive BH
binary at a given separation \citep[\textit{e.g.},][]{HKM09,
PG1302MNRAS:2015a}, compared to the probability of observing such a
disruption, computed here. That is, what is more improbable: the
chance of finding a TDF from such a large BH, out of a sample of
$\mathcal{O}(10)$; or, the chance of finding a super massive BH binary with
the required mass ratio in a range of orbital separations small enough to not
alter the BH mass measurement, but large enough to not alter the TDF light
curve.

For a super massive BH binary with an orbital period on order the TDF
timescale, the TDF lightcurve may hold clues to its binary origin, or perhaps
it would not be recognizable at all \citep{HayasakiLoeb:2016,
Couhglin_MBHBTDE_1+2017, Liu_MBHBTDE_cand+2014, Vigneron+2018}. It would be
interesting to consider at what level a super massive BH binary population
could effect the BH mass-dependent TDF rate
\citep[\textit{e.g.},][]{FragioneLeigh:2018} and include this as a factor into
the model presented here.

Interpretation of the BH mass dependent TDF rate could also be
affected by the presence of a different mass-radius relation than adopted
here, \textit{e.g.}, involving the disruption of a non-main sequence star such
as a red giant. This, however, would likely cause a discernibly different TDF
\citep[\textit{e.g.}][]{MacleodSpoon+2013}. While red giants and other 
non-main sequence stars make up only a small fraction ($\lsim10\%$) of the stellar
population, a large enough TDF sample may begin to identify non-main sequence
star disruptions. In that case the analysis presented here, but adapted to the
different stellar species, could serve as an additional constraint on the PDMF
probed through  main-sequence disruptions.

In summary, we have demonstrated a novel technique for constraining the PDMFs
of nuclear star clusters from tidal disruptions of stars by massive BHs
and also presented the most up to date version of the differential TDF
rate per BH mass using the BH masses of TDFs from both V18 and M18. We
conclude with a number of additional caveats and topics for future
exploration.

\begin{itemize}
\item \textit{BH spin:}
The range of BH spins could generate an intrinsic scatter in the TDF rate. (see also v18 for discussion of spin inference from the BH mass dependent TDF rate).
\item \textit{Metallicity range:}
The mass to radius relation the we use for main sequence stars, derived from the fit of \citep{Tout+1996}, breaks down for metallicities above $\sim3 Z_{\odot}$. However, there is observational evidence for higher metallicity environments in the Galactic Centre \citep{TDo+2018}. 
Note further that non-main sequence stars, or stars with pericentre just larger than the tidal disruption radius, may have a different mass-to radius relation entirely \citep[\textit{e.g.}][]{MacleodSpoon+2013}, regardless of metallicity.
\item{
\textit{Redshift dependence:} 
One could introduce a differential TDF with redshift as well as BH mass by binning the TDFs in redshift. With enough TDFs (approximately the number of redshift bins times the numbers quoted in each analysis of this work), this would allow one to track the redshift evolution of the stellar mass function that might be expected from, \textit{e.g.}, redshift dependent metallicity evolution \citep{Belczynski+2016_Zevoz}.}
\item \textit{TDF rate cutoff at low BH mass:}
If the TDF luminosity is limited by the Eddington luminosity of the BH
\citep{DeColle+2012}, or if the local BH mass function drops off due to the
absence of intermediate mass BHs ($10^4-10^5 \Msun$), then the BH-mass
dependent TDF rate should drop off also at small BH masses. An 
Eddington-limited observational bias could be included in future analyses of the TDF
rate in order to investigate the low end of the BH mass function (see also v18 
for discussion of the low end of the BH mass function in this context).
\item \textit{Dependence on the BH mass function:}
As a related point, we have used a single BH mass function and not allowed the values of the parameters determining this to vary. With the many future TDFs expected at lower BH masses, future data could also constrain the local BH mass function in a higher dimensional parameter space version of the analysis presented here. 
\item{\textit{Form of the model PDMF:} 
For clarity, we have presented our analysis for only one class of PDMF model,
namely a broken power law. Future analyses could also examine the ability to
select between model classes (\textit{e.g.} power law vs. exponential cutoff),
or introduce a more general PDMF model.}
\item \textit{Stellar mass preference for disruption?:}
In using the TDF rate to constrain the PDMF, we assume that there is no
preference for a given mass star to be on a disrupting orbit. If there is
preference for a subset of stellar masses to be perturbed onto centrophilic
orbits, then the work here is constraining the mass function only of those stars.
\item \textit{Generation of BH mass dependent TDF rate:}
When creating the binned data in Figure~\ref{Fig:3M}, we opted to choose bin sizes that reproduce the data from V18. However, the bin choice is somewhat arbitrary. This will likely not be an issue once the number of TDFs increases in the LSST era, but if more information is to be gleaned from the existing data, the effects of bin size should be explored. Alternatively, one might abandon binning entirely and fit to a cumulative distribution function rather than the probability distribution function used here.
\end{itemize}

The exploration of each of the above items would prove useful for developing a
powerful new tool for probing the unique stellar environments surrounding super massive
BHs in the centers of galactic nuclei.

\section*{Acknowledgements}
The authors thank Brenna Mockler, Ben Johnson, Rebbecca Oppenheimer, Will
Farr, Josh Spiegel, and Sjoert van Velzen for useful comments, discussions,
and information. DJD thanks Norm Murray for his talk at the Black Hole
Initiative that inspired this work. The authors thank the anonymous referee
for an insightful review that improved the quality of this manuscript.
Financial support was provided from NASA through Einstein Postdoctoral
Fellowship award number PF6-170151 (DJD). This work was supported in part by
the Black Hole Initiative, which is funded by a grant from the John Templeton
Foundation.

\appendix
\section{Analytic Expression for solar-metallicity, Salpeter mass function models}

Setting $Z=Z_{\odot}$, and using only a single power law (Salpeter) mass
function, the probability $\mathcal{P}$ (Eq. \ref{Eq:Bayes}) can be written
analytically in a simple expression. We include it here for use in cursory
estimates of the BH mass dependent TDF rate

For $Z=Z_{\odot}$, the main-sequence mass-radius relation is approximated as,
\begin{equation}
    r_* = \left(\frac{M_*}{\Msun}\right)^{\xi} \Rsun, \quad \mbox{where} \
    \begin{cases} 
    \xi=0.85 & M \leq \Msun \nonumber \\
   \xi=0.6 & M > \Msun
    \end{cases}.
\label{AEq:MR}
\end{equation}
Then the critical mass is
\begin{equation}
M_{\rm{crit}}(\xi) \equiv \left[\left( \frac{\left<r_{\IBCO}\right>_{\spin}}{M_{\BHcirc}} \frac{\Msun}{\Rsun} \right)^3 M^{(3\xi -3)}_{\odot} M^2_{\BHcirc} \right]^{1/(3 \xi - 1)},
\label{AEq:Mcrit1}
\end{equation}
which we write as,
\begin{equation}
M^*_{\rm{crit}} = \rm{Max}\left\{ M_{\rm{crit}}(0.85), M_{\rm{crit}}(0.6)\right\},
\label{AEq:Mcrit}
\end{equation}
to take into account the change in mass-radius relation on the main sequence.

Assuming a PDMF that is the result of a power law (\textit{e.g.}, Salpeter)
initial stellar mass function \citep{MagTrem:1999},
\begin{equation}
\frac{dN_*}{dM_*} \propto \left( \frac{M_*}{\Msun}\right)^{\xi} H(M_* - M_{\min}) [1 - H(M_* - M_{\max})],
\label{AEq:SMF}
\end{equation}
with lower and upper stellar mass cutoffs $M_{\min}$ and $M_{\max}$ enforced by the Heaviside function $H$.

Then carrying out the integrals in Eq. (\ref{Eq:Pdef}), we find that in this simplified, but still useful case, the probability of a disruption, given the BH mass, is given by the analytic expression,
\begin{eqnarray}
&P( \rm{TDF} | M_{\BHcirc} ) = \frac{\left\{1 - H\left(M^*_{\crit} - M_{\max}\right)  \right\} }{1+\xi}  \times \\  \nonumber 
&\left\{ \left[M^{1+\xi}_{\min} - M^{1+\xi}_{\crit}\right] \left[M^{1+\xi}_{\max} - M^{1+\xi}_{\min}  + H\left(M^*_{\crit} - M_{\min}\right)  \right] \right\} ,
\label{AEq:PdefAnl}
\end{eqnarray}
which is substituted into Eq. (\ref{Eq:Bayes}) to get $\mathcal{P}$.

\bibliographystyle{mnras}
\bibliography{refs}

\begin{thebibliography}{}
\makeatletter
\relax
\def\mn@urlcharsother{\let\do\@makeother \do\$\do\&\do\#\do\^\do\_\do\%\do\~}
\def\mn@doi{\begingroup\mn@urlcharsother \@ifnextchar [ {\mn@doi@}
  {\mn@doi@[]}}
\def\mn@doi@[#1]#2{\def\@tempa{#1}\ifx\@tempa\@empty \href
  {http://dx.doi.org/#2} {doi:#2}\else \href {http://dx.doi.org/#2} {#1}\fi
  \endgroup}
\def\mn@eprint#1#2{\mn@eprint@#1:#2::\@nil}
\def\mn@eprint@arXiv#1{\href {http://arxiv.org/abs/#1} {{\tt arXiv:#1}}}
\def\mn@eprint@dblp#1{\href {http://dblp.uni-trier.de/rec/bibtex/#1.xml}
  {dblp:#1}}
\def\mn@eprint@#1:#2:#3:#4\@nil{\def\@tempa {#1}\def\@tempb {#2}\def\@tempc
  {#3}\ifx \@tempc \@empty \let \@tempc \@tempb \let \@tempb \@tempa \fi \ifx
  \@tempb \@empty \def\@tempb {arXiv}\fi \@ifundefined
  {mn@eprint@\@tempb}{\@tempb:\@tempc}{\expandafter \expandafter \csname
  mn@eprint@\@tempb\endcsname \expandafter{\@tempc}}}

\bibitem[\protect\citeauthoryear{{Arcavi} et~al.,}{{Arcavi}
  et~al.}{2014}]{Arcavi+2014}
{Arcavi} I.,  et~al., 2014, \mn@doi [\apj] {10.1088/0004-637X/793/1/38}, \href
  {http://adsabs.harvard.edu/abs/2014ApJ...793...38A} {793, 38}

\bibitem[\protect\citeauthoryear{{Bardeen}, {Press}  \& {Teukolsky}}{{Bardeen}
  et~al.}{1972}]{Bardeen72}
{Bardeen} J.~M.,  {Press} W.~H.,   {Teukolsky} S.~A.,  1972, \mn@doi [\apj]
  {10.1086/151796}, \href {http://adsabs.harvard.edu/abs/1972ApJ...178..347B}
  {178, 347}

\bibitem[\protect\citeauthoryear{{Bartko} et~al.,}{{Bartko}
  et~al.}{2010}]{Bartko_GCSMF+2010}
{Bartko} H.,  et~al., 2010, \mn@doi [\apj] {10.1088/0004-637X/708/1/834}, \href
  {http://adsabs.harvard.edu/abs/2010ApJ...708..834B} {708, 834}

\bibitem[\protect\citeauthoryear{{Belczynski}, {Holz}, {Bulik}  \&
  {O'Shaughnessy}}{{Belczynski} et~al.}{2016}]{Belczynski+2016_Zevoz}
{Belczynski} K.,  {Holz} D.~E.,  {Bulik} T.,   {O'Shaughnessy} R.,  2016,
  \mn@doi [\nat] {10.1038/nature18322}, \href
  {http://adsabs.harvard.edu/abs/2016Natur.534..512B} {534, 512}

\bibitem[\protect\citeauthoryear{{Bloom} et~al.,}{{Bloom}
  et~al.}{2011}]{Bloom+2011}
{Bloom} J.~S.,  et~al., 2011, \mn@doi [Science] {10.1126/science.1207150},
  \href {http://adsabs.harvard.edu/abs/2011Sci...333..203B} {333, 203}

\bibitem[\protect\citeauthoryear{{Brockamp}, {Baumgardt}  \&
  {Kroupa}}{{Brockamp} et~al.}{2011}]{Brockamp+2011}
{Brockamp} M.,  {Baumgardt} H.,   {Kroupa} P.,  2011, \mn@doi [\mnras]
  {10.1111/j.1365-2966.2011.19580.x}, \href
  {http://adsabs.harvard.edu/abs/2011MNRAS.418.1308B} {418, 1308}

\bibitem[\protect\citeauthoryear{{Chornock} et~al.,}{{Chornock}
  et~al.}{2014}]{Chornock+2014}
{Chornock} R.,  et~al., 2014, \mn@doi [\apj] {10.1088/0004-637X/780/1/44},
  \href {http://adsabs.harvard.edu/abs/2014ApJ...780...44C} {780, 44}

\bibitem[\protect\citeauthoryear{{Coughlin} \& {Armitage}}{{Coughlin} \&
  {Armitage}}{2018}]{Coughlin15lh+2018}
{Coughlin} E.~R.,  {Armitage} P.~J.,  2018, \mn@doi [\mnras]
  {10.1093/mnras/stx3039}, \href
  {http://adsabs.harvard.edu/abs/2018MNRAS.474.3857C} {474, 3857}

\bibitem[\protect\citeauthoryear{{Coughlin}, {Armitage}, {Nixon}  \&
  {Begelman}}{{Coughlin} et~al.}{2017}]{Couhglin_MBHBTDE_1+2017}
{Coughlin} E.~R.,  {Armitage} P.~J.,  {Nixon} C.,   {Begelman} M.~C.,  2017,
  \mn@doi [\mnras] {10.1093/mnras/stw2913}, \href
  {http://adsabs.harvard.edu/abs/2017MNRAS.465.3840C} {465, 3840}

\bibitem[\protect\citeauthoryear{{D'Orazio} \& {Levin}}{{D'Orazio} \&
  {Levin}}{2013}]{DL:2013}
{D'Orazio} D.~J.,  {Levin} J.,  2013, \mn@doi [\prd]
  {10.1103/PhysRevD.88.064059}, \href
  {http://adsabs.harvard.edu/abs/2013PhRvD..88f4059D} {88, 064059}

\bibitem[\protect\citeauthoryear{{D'Orazio}, {Haiman}, {Duffell}, {Farris}  \&
  {MacFadyen}}{{D'Orazio} et~al.}{2015}]{PG1302MNRAS:2015a}
{D'Orazio} D.~J.,  {Haiman} Z.,  {Duffell} P.,  {Farris} B.~D.,   {MacFadyen}
  A.~I.,  2015, \mn@doi [\mnras] {10.1093/mnras/stv1457}, \href
  {http://adsabs.harvard.edu/abs/2015MNRAS.452.2540D} {452, 2540}

\bibitem[\protect\citeauthoryear{{D'Orazio}, {Levin}, {Murray}  \&
  {Price}}{{D'Orazio} et~al.}{2016}]{DL:2016}
{D'Orazio} D.~J.,  {Levin} J.,  {Murray} N.~W.,   {Price} L.,  2016, preprint,
  \href {http://adsabs.harvard.edu/abs/2016arXiv160100017D} {} (\mn@eprint
  {arXiv} {1601.00017})

\bibitem[\protect\citeauthoryear{{De Colle}, {Guillochon}, {Naiman}  \&
  {Ramirez-Ruiz}}{{De Colle} et~al.}{2012}]{DeColle+2012}
{De Colle} F.,  {Guillochon} J.,  {Naiman} J.,   {Ramirez-Ruiz} E.,  2012,
  \mn@doi [\apj] {10.1088/0004-637X/760/2/103}, \href
  {http://adsabs.harvard.edu/abs/2012ApJ...760..103D} {760, 103}

\bibitem[\protect\citeauthoryear{{Do}, {Kerzendorf}, {Konopacky}, {Marcinik},
  {Ghez}, {Lu}  \& {Morris}}{{Do} et~al.}{2018}]{TDo+2018}
{Do} T.,  {Kerzendorf} W.,  {Konopacky} Q.,  {Marcinik} J.~M.,  {Ghez} A.,
  {Lu} J.~R.,   {Morris} M.~R.,  2018, \mn@doi [\apjl]
  {10.3847/2041-8213/aaaec3}, \href
  {http://adsabs.harvard.edu/abs/2018ApJ...855L...5D} {855, L5}

\bibitem[\protect\citeauthoryear{{Dong} et~al.,}{{Dong}
  et~al.}{2016}]{Dong15lh+2016}
{Dong} S.,  et~al., 2016, \mn@doi [Science] {10.1126/science.aac9613}, \href
  {http://adsabs.harvard.edu/abs/2016Sci...351..257D} {351, 257}

\bibitem[\protect\citeauthoryear{{Fragione} \& {Leigh}}{{Fragione} \&
  {Leigh}}{2018}]{FragioneLeigh:2018}
{Fragione} G.,  {Leigh} N.,  2018, \mn@doi [\mnras] {10.1093/mnras/sty1600},
  \href {http://adsabs.harvard.edu/abs/2018MNRAS.479.3181F} {479, 3181}

\bibitem[\protect\citeauthoryear{{Gezari} et~al.,}{{Gezari}
  et~al.}{2006}]{Gezari+2006}
{Gezari} S.,  et~al., 2006, \mn@doi [\apjl] {10.1086/509918}, \href
  {http://adsabs.harvard.edu/abs/2006ApJ...653L..25G} {653, L25}

\bibitem[\protect\citeauthoryear{{Gezari} et~al.,}{{Gezari}
  et~al.}{2009}]{Gezari+2009}
{Gezari} S.,  et~al., 2009, \mn@doi [\apj] {10.1088/0004-637X/698/2/1367},
  \href {http://adsabs.harvard.edu/abs/2009ApJ...698.1367G} {698, 1367}

\bibitem[\protect\citeauthoryear{{Gezari} et~al.,}{{Gezari}
  et~al.}{2012}]{Gezari+2012}
{Gezari} S.,  et~al., 2012, \mn@doi [\nat] {10.1038/nature10990}, \href
  {http://adsabs.harvard.edu/abs/2012Natur.485..217G} {485, 217}

\bibitem[\protect\citeauthoryear{Haiman, Kocsis  \& Menou}{Haiman
  et~al.}{2009}]{HKM09}
Haiman Z.,  Kocsis B.,   Menou K.,  2009, \mn@doi [\apj]
  {10.1088/0004-637X/700/2/1952}, 700, 1952

\bibitem[\protect\citeauthoryear{{Hayasaki} \& {Loeb}}{{Hayasaki} \&
  {Loeb}}{2016}]{HayasakiLoeb:2016}
{Hayasaki} K.,  {Loeb} A.,  2016, \mn@doi [Scientific Reports]
  {10.1038/srep35629}, \href
  {http://adsabs.harvard.edu/abs/2016NatSR...635629H} {6, 35629}

\bibitem[\protect\citeauthoryear{{Higson}, {Handley}, {Hobson}  \&
  {Lasenby}}{{Higson} et~al.}{2017}]{Higson:2017}
{Higson} E.,  {Handley} W.,  {Hobson} M.,   {Lasenby} A.,  2017, preprint,
  \href {http://adsabs.harvard.edu/abs/2017arXiv170403459H} {} (\mn@eprint
  {arXiv} {1704.03459})

\bibitem[\protect\citeauthoryear{Hills}{Hills}{1975}]{Hills:1975}
Hills J.~G.,  1975, \nat, 254, 295

\bibitem[\protect\citeauthoryear{{Hod}}{{Hod}}{2017}]{Hod:2017}
{Hod} S.,  2017, preprint, \href
  {http://adsabs.harvard.edu/abs/2017arXiv170705680H} {} (\mn@eprint {arXiv}
  {1707.05680})

\bibitem[\protect\citeauthoryear{{Holoien} et~al.,}{{Holoien}
  et~al.}{2014}]{Holoien+2014}
{Holoien} T.~W.-S.,  et~al., 2014, \mn@doi [\mnras] {10.1093/mnras/stu1922},
  \href {http://adsabs.harvard.edu/abs/2014MNRAS.445.3263H} {445, 3263}

\bibitem[\protect\citeauthoryear{{Hung} et~al.,}{{Hung}
  et~al.}{2017}]{SiftSaph:2017}
{Hung} T.,  et~al., 2017, preprint, \href
  {http://adsabs.harvard.edu/abs/2017arXiv171204936H} {} (\mn@eprint {arXiv}
  {1712.04936})

\bibitem[\protect\citeauthoryear{{Ivezic} \& et al.}{{Ivezic} \&
  et~al.}{2008}]{LSST}
{Ivezic} Z.,  et al. 2008, preprint, \href
  {http://adsabs.harvard.edu/abs/2008arXiv0805.2366I} {} (\mn@eprint {arXiv}
  {0805.2366})

\bibitem[\protect\citeauthoryear{{Kochanek}}{{Kochanek}}{2016}]{Kochanek:2016}
{Kochanek} C.~S.,  2016, \mn@doi [\mnras] {10.1093/mnras/stw1290}, \href
  {http://adsabs.harvard.edu/abs/2016MNRAS.461..371K} {461, 371}

\bibitem[\protect\citeauthoryear{{Kr{\"u}hler} et~al.,}{{Kr{\"u}hler}
  et~al.}{2018}]{Kruhler+2018}
{Kr{\"u}hler} T.,  et~al., 2018, \mn@doi [\aap] {10.1051/0004-6361/201731773},
  \href {http://adsabs.harvard.edu/abs/2018A%26A...610A..14K} {610, A14}

\bibitem[\protect\citeauthoryear{{Leloudas} et~al.,}{{Leloudas}
  et~al.}{2016}]{Leloudas+2016}
{Leloudas} G.,  et~al., 2016, \mn@doi [Nature Astronomy]
  {10.1038/s41550-016-0002}, \href
  {http://adsabs.harvard.edu/abs/2016NatAs...1E...2L} {1, 0002}

\bibitem[\protect\citeauthoryear{{Levin} \& {Perez-Giz}}{{Levin} \&
  {Perez-Giz}}{2009}]{LevinGabe:2009}
{Levin} J.,  {Perez-Giz} G.,  2009, \mn@doi [\prd]
  {10.1103/PhysRevD.79.124013}, \href
  {http://adsabs.harvard.edu/abs/2009PhRvD..79l4013L} {79, 124013}

\bibitem[\protect\citeauthoryear{{Liu}, {Li}  \& {Komossa}}{{Liu}
  et~al.}{2014}]{Liu_MBHBTDE_cand+2014}
{Liu} F.~K.,  {Li} S.,   {Komossa} S.,  2014, \mn@doi [\apj]
  {10.1088/0004-637X/786/2/103}, \href
  {http://adsabs.harvard.edu/abs/2014ApJ...786..103L} {786, 103}

\bibitem[\protect\citeauthoryear{{Lu}, {Kumar}  \& {Narayan}}{{Lu}
  et~al.}{2017}]{Lu:2017a}
{Lu} W.,  {Kumar} P.,   {Narayan} R.,  2017, \mnras, 468, 910

\bibitem[\protect\citeauthoryear{{MacLeod}, {Ramirez-Ruiz}, {Grady}  \&
  {Guillochon}}{{MacLeod} et~al.}{2013}]{MacleodSpoon+2013}
{MacLeod} M.,  {Ramirez-Ruiz} E.,  {Grady} S.,   {Guillochon} J.,  2013,
  \mn@doi [\apj] {10.1088/0004-637X/777/2/133}, \href
  {http://adsabs.harvard.edu/abs/2013ApJ...777..133M} {777, 133}

\bibitem[\protect\citeauthoryear{{Magorrian} \& {Tremaine}}{{Magorrian} \&
  {Tremaine}}{1999}]{MagTrem:1999}
{Magorrian} J.,  {Tremaine} S.,  1999, \mn@doi [\mnras]
  {10.1046/j.1365-8711.1999.02853.x}, \href
  {http://adsabs.harvard.edu/abs/1999MNRAS.309..447M} {309, 447}

\bibitem[\protect\citeauthoryear{{Margutti} et~al.,}{{Margutti}
  et~al.}{2017}]{Margutti+2017}
{Margutti} R.,  et~al., 2017, \mn@doi [\apj] {10.3847/1538-4357/836/1/25},
  \href {http://adsabs.harvard.edu/abs/2017ApJ...836...25M} {836, 25}

\bibitem[\protect\citeauthoryear{{Metzger}, {Margalit}, {Kasen}  \&
  {Quataert}}{{Metzger} et~al.}{2015}]{Metzger+2015}
{Metzger} B.~D.,  {Margalit} B.,  {Kasen} D.,   {Quataert} E.,  2015, \mn@doi
  [\mnras] {10.1093/mnras/stv2224}, \href
  {http://adsabs.harvard.edu/abs/2015MNRAS.454.3311M} {454, 3311}

\bibitem[\protect\citeauthoryear{{Mockler}, {Guillochon}  \&
  {Ramirez-Ruiz}}{{Mockler} et~al.}{2018}]{MocklerJames+2018}
{Mockler} B.,  {Guillochon} J.,   {Ramirez-Ruiz} E.,  2018, preprint, \href
  {http://adsabs.harvard.edu/abs/2018arXiv180108221M} {} (\mn@eprint {arXiv}
  {1801.08221})

\bibitem[\protect\citeauthoryear{{Rees}}{{Rees}}{1988}]{Rees:1988}
{Rees} M.~J.,  1988, \mn@doi [\nat] {10.1038/333523a0}, \href
  {http://adsabs.harvard.edu/abs/1988Natur.333..523R} {333, 523}

\bibitem[\protect\citeauthoryear{{Shankar}, {Salucci}, {Granato}, {De Zotti}
  \& {Danese}}{{Shankar} et~al.}{2004}]{Shankar+2004}
{Shankar} F.,  {Salucci} P.,  {Granato} G.~L.,  {De Zotti} G.,   {Danese} L.,
  2004, \mn@doi [\mnras] {10.1111/j.1365-2966.2004.08261.x}, \href
  {http://adsabs.harvard.edu/abs/2004MNRAS.354.1020S} {354, 1020}

\bibitem[\protect\citeauthoryear{{Stone} \& {Metzger}}{{Stone} \&
  {Metzger}}{2016}]{StoneMetzger:2016}
{Stone} N.~C.,  {Metzger} B.~D.,  2016, \mn@doi [\mnras]
  {10.1093/mnras/stv2281}, \href
  {http://adsabs.harvard.edu/abs/2016MNRAS.455..859S} {455, 859}

\bibitem[\protect\citeauthoryear{{Sukhbold} \& {Woosley}}{{Sukhbold} \&
  {Woosley}}{2016}]{SukhWoos+2016}
{Sukhbold} T.,  {Woosley} S.~E.,  2016, \mn@doi [\apjl]
  {10.3847/2041-8205/820/2/L38}, \href
  {http://adsabs.harvard.edu/abs/2016ApJ...820L..38S} {820, L38}

\bibitem[\protect\citeauthoryear{{Tadhunter}, {Spence}, {Rose}, {Mullaney}  \&
  {Crowther}}{{Tadhunter} et~al.}{2017}]{Tadhunter+2017}
{Tadhunter} C.,  {Spence} R.,  {Rose} M.,  {Mullaney} J.,   {Crowther} P.,
  2017, \mn@doi [Nature Astronomy] {10.1038/s41550-017-0061}, \href
  {http://adsabs.harvard.edu/abs/2017NatAs...1E..61T} {1, 0061}

\bibitem[\protect\citeauthoryear{{Tout}, {Pols}, {Eggleton}  \& {Han}}{{Tout}
  et~al.}{1996}]{Tout+1996}
{Tout} C.~A.,  {Pols} O.~R.,  {Eggleton} P.~P.,   {Han} Z.,  1996, \mn@doi
  [\mnras] {10.1093/mnras/281.1.257}, \href
  {http://adsabs.harvard.edu/abs/1996MNRAS.281..257T} {281, 257}

\bibitem[\protect\citeauthoryear{{Vigneron}, {Lodato}  \&
  {Guidarelli}}{{Vigneron} et~al.}{2018}]{Vigneron+2018}
{Vigneron} Q.,  {Lodato} G.,   {Guidarelli} A.,  2018, \mn@doi [\mnras]
  {10.1093/mnras/sty585}, \href
  {http://adsabs.harvard.edu/abs/2018MNRAS.476.5312V} {476, 5312}

\bibitem[\protect\citeauthoryear{{Vink{\'o}} et~al.,}{{Vink{\'o}}
  et~al.}{2015}]{Vinko+2015}
{Vink{\'o}} J.,  et~al., 2015, \mn@doi [\apj] {10.1088/0004-637X/798/1/12},
  \href {http://adsabs.harvard.edu/abs/2015ApJ...798...12V} {798, 12}

\bibitem[\protect\citeauthoryear{{Wang} \& {Merritt}}{{Wang} \&
  {Merritt}}{2004}]{WangMerritt:2004}
{Wang} J.,  {Merritt} D.,  2004, \mn@doi [\apj] {10.1086/379767}, \href
  {http://adsabs.harvard.edu/abs/2004ApJ...600..149W} {600, 149}

\bibitem[\protect\citeauthoryear{{Will}}{{Will}}{2012}]{Will:2012}
{Will} C.~M.,  2012, \mn@doi [Classical and Quantum Gravity]
  {10.1088/0264-9381/29/21/217001}, \href
  {http://adsabs.harvard.edu/abs/2012CQGra..29u7001W} {29, 217001}

\bibitem[\protect\citeauthoryear{{van Velzen}}{{van
  Velzen}}{2018}]{vanVelzen:2018}
{van Velzen} S.,  2018, \mn@doi [\apj] {10.3847/1538-4357/aa998e}, \href
  {http://adsabs.harvard.edu/abs/2018ApJ...852...72V} {852, 72}

\bibitem[\protect\citeauthoryear{{van Velzen} et~al.,}{{van Velzen}
  et~al.}{2011}]{vanVelzen+2011}
{van Velzen} S.,  et~al., 2011, \mn@doi [\apj] {10.1088/0004-637X/741/2/73},
  \href {http://adsabs.harvard.edu/abs/2011ApJ...741...73V} {741, 73}

\makeatother
\end{thebibliography}
\end{document}